\documentclass[10pt,a4paper,english]{article}
\usepackage{graphicx}

\begin{document}

\title{Derivation of Gell-Mann-Nishijima formula from the electromagnetic
field modes of a hadron}

\author{{\small Huai-yang Cui}\\
 {\small Department of Physics, Beihang University}\\
 {\small Beijing, 100191, China, E-mail: hycui@buaa.edu.cn}}

\date{{\small \today}}
\maketitle
\begin{abstract}
When an electron probes another elementary particle Q, the wave function
of the electron can be separated into two independent parts, the first
part represents the electronic motion, the second part represents
the electromagnetic field mode around the particle Q. In analogy with
optical modes $TEM_{nlm}$ for a laser resonator, when the electromagnetic
field around the particle Q forms into a mode, the quantum numbers
of the mode satisfy the Gell-Mann-Nishijima formula, these quantum
numbers are recognized as the charge number, baryon number and strangeness
number. The modes are used as a visual model to understand the abstract
baryon number and strangeness number of hadrons.

 \\

\end{abstract}

\section{Introduction}

The Gell-Mann-Nishijima formula relates the baryon number $B$, the
strangeness number $S$, the isospin $I_{3}$ of hadrons to the charge
number $Q$. It was originally given by Tadao Nakano and Kazuhiko
Nishijima in 1953\cite{T_Nakano,K_Nishijima}, Murray Gell-Mann proposed
the formula independently in 1956\cite{M_Gell_Mann}. The modern version
of the formula relates all flavor quantum numbers with the baryon
number and the electric charge. We can understand the charge number
$Q$ through the Coulomb's interaction, but failed to understand the
baryon number $B$ and the strangeness number $S$ with reality, so
$B$, $S$ and $I_{3}$ are abstract. Since $B$ and $S$ appear in
the formula with the charge number $Q$, they must relate to the electromagnetic
interaction in some way, this clue guide us to study the possible
electromagnetic field modes around a hadron, in analogy with optical
modes $TEM_{nlm}$ for a laser resonator. As a result, it is found
that the Gell-Mann-Nishijima formula is the critical condition of
the electromagnetic field modes around a hadron. It is important to
realize that the electromagnetic field modes around a hadron provide
us a visual model to understand the abstract baryon number $B$ and
strangeness number $S$ in reality.

\section{Electromagnetic field around a particle}

\begin{figure}[htb]
 \centering \includegraphics[bb=160bp 590bp 420bp 760bp,clip]{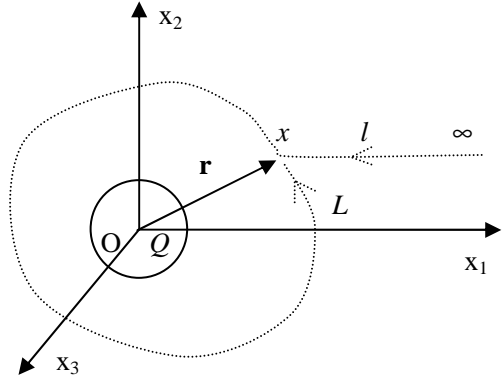}
\caption{The calculation of the wave function around a particle by path integral.}

\label{Fig1} 
\end{figure}

Consider an electron probing an elementary particle $Q$, the wave
function of the electron is given by\cite{H_Y_Cui}

\begin{equation}
(p_{\mu}-eA_{\mu})\psi=-i\hbar\partial_{\mu}\psi\label{gn1}\end{equation}
 \begin{equation}
\psi(x)=exp[\frac{i}{\hbar}\int_{l}(p_{\mu}-eA_{\mu})dx_{\mu}]\label{gn2}\end{equation}
 Where $A_{\mu}$ is the 4-vector potential around the particle Q.
Frequently, $p_{\mu}$ in Eq.(\ref{gn1}) is understood as momentum
operator of the electron, but here we regard $p_{\mu}$ as momentum
itself as shown in Eq.(\ref{gn2}) for calculating the wave function.
Because of path independence, $l$ is an arbitrary path from the initial
point $x_{0}$ to the observed point $x$, we can choose two path
$l$ and $l+L$ to calculate the wave function, as shown in Figure
\ref{Fig1}, $L$ is a closed loop linking the observed point $x$
around the particle $Q$, then

\begin{eqnarray}
\psi(x)_{l} & = & \psi(x)_{l+L}\label{gn3}\\
exp[\frac{i}{\hbar}\int_{l}(p_{\mu}-eA_{\mu})dx_{\mu}] & = & exp[\frac{i}{\hbar}\int_{l+L}(p_{\mu}-eA_{\mu})dx_{\mu}]\label{gn4}\end{eqnarray}
 We obtain a quantization equation:

\begin{equation}
\frac{1}{\hbar}\int_{L}(p_{\mu}-eA_{\mu})dx_{\mu}=2\pi n\quad n=0,\pm1,\pm2,...\label{gn5}\end{equation}

The second part of the loop integral, $\frac{1}{\hbar}\int_{L}(-eA_{\mu})dx_{\mu}$,
is independent from the first part $\frac{1}{\hbar}\int_{L}(p_{\mu})dx_{\mu}$,
no mater how much the electron energy $\varepsilon$ is ($\varepsilon=1MeV$
or $\varepsilon=500MeV$), it always hold the quantization equation,
so we have

\begin{equation}
\frac{1}{\hbar}\int_{L}(-eA_{\mu})dx_{\mu}=2\pi n'\quad n'=0,\pm1,\pm2,...\label{gn6}\end{equation}
 This part represents the properties of the particle $Q$, and defining\begin{equation}
\xi(x)=exp[\frac{i}{\hbar}\int_{l}(-eA_{\mu})dx_{\mu}]\label{gn7}\end{equation}
 We call $\xi(x)$ as the potential-wave-function of the particle
$Q$, Eq.(\ref{gn6}) is the quantization equation of the potential-wave-function.

\section{Magnetism of particle}

If the particle $Q$ has magnetic field around it without electric
field, $\mathbf{E}=0$, then \begin{equation}
A_{4}(x)=\frac{i}{c}V(x)=\frac{i}{c}\int_{x}^{\infty}\mathbf{E}\cdot d\mathbf{l}=0\label{gn8}\end{equation}
 The quantized Eq.(\ref{gn6}) reduces to \begin{equation}
\frac{-e}{\hbar}\int_{L}\mathbf{A}\cdot d\mathbf{l}=2\pi n\quad n=0,\pm1,\pm2,...\label{gn9}\end{equation}
 Because of magnetic field $\mathbf{B}=\nabla\times\mathbf{A}$, we
find that the magnetic flux $\Phi_{m}$through the closed loop $L$
is quantized by \begin{equation}
\Phi_{m}=\int_{S}\mathbf{B}\cdot d\mathbf{s}=\int_{S}\mathbf{(\nabla\times\mathbf{A})}\cdot d\mathbf{s}=\oint_{L}\mathbf{A}\cdot d\mathbf{l}=-2\pi n\frac{\hbar}{e}\label{gn10}\end{equation}
 It is important to note that the loop shape $L$ or surface area
$S$ are arbitrary because of path independence for the potential-wave-function,
the magnetic flux must concentrate into the particle interior as a
tiny solenoid, the particle position is a singularity in mathematics.
Experiments have made numerous measurements on the magnetic moments
of baryons, but lacking attentions to magnetic fluxes of baryons.

\section{Gell-Mann-Nishijima formula for hadrons}

\begin{figure}[htb]
 \centering \includegraphics[bb=130bp 560bp 400bp 750bp,clip]{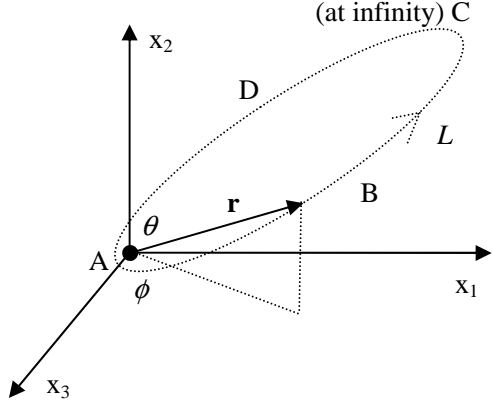}
\caption{The big closed loop ABCDA around a particle.}

\label{Fig2} 
\end{figure}

In order to describe the 4D electromagnetic field around the particle
$Q$, we choose a big closed loop $L$ as shown in Figure \ref{Fig2},
saying A-B-C-D-A, where A contains the particle at the origin and
C at the infinity, from Eq.(\ref{gn6}) we have \begin{eqnarray}
\frac{-e}{\hbar}\oint_{L}A_{r}\cdot dr+\frac{-e}{\hbar}\oint_{L}A_{\theta}\cdot rd\theta+\frac{-e}{\hbar}\oint_{L}A_{\phi}\cdot r\sin\theta d\phi+\frac{-e}{\hbar}\oint_{L}A_{4}\cdot d(ict) & = & 2\pi n\label{gn11}\\
n & = & 0,\pm1,\pm2,...\nonumber \end{eqnarray}

The big closed loop $L$ makes that the angle $\theta$ varies from
$0$ to $2\pi$ and the angle $\phi$ varies from $0$ to $2\pi$,
the radical distance $r$ varies from 0 to $R=\infty$ then returns
to $0$, time t varies from 0 to $T=\infty$ then returns to $0$.
Note that $L$ is not the path of the probing electron, is a calculation
path for wave function.

\begin{figure}[htb]
 \centering \includegraphics[bb=115bp 580bp 480bp 740bp,clip]{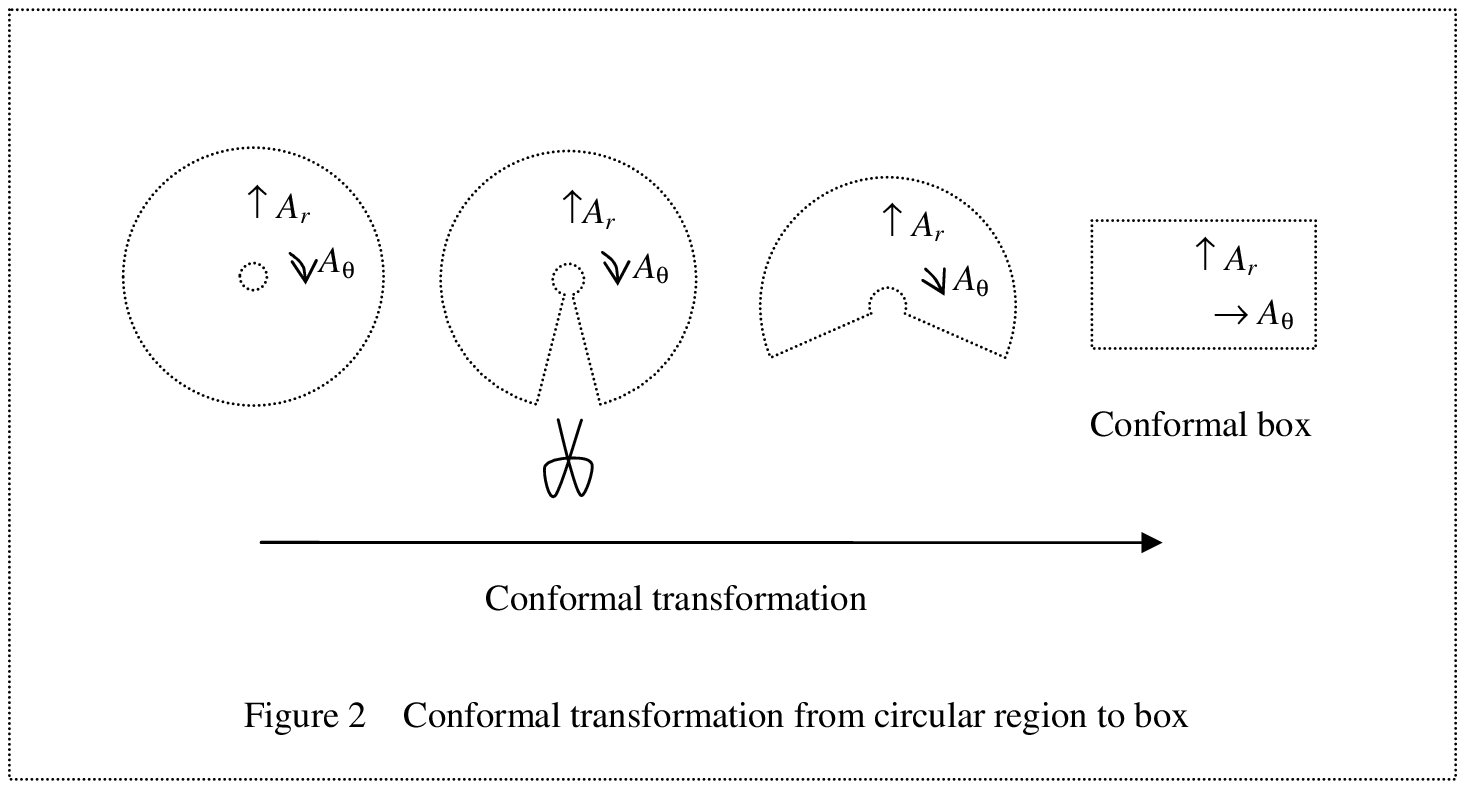}
\caption{A conformal transformation from the circular region to the conformal
box.}

\label{Fig3} 
\end{figure}

We can imagine a conformal transformation for the electromagnetic
field around the nucleus, as shown in Figure \ref{Fig3}, from 4D
$A$ in the circular region to 4D $A$ in the conformal box. (conformal
theory in mathematics is a classical branch). We use this conformal
box to emphasize that we must fairly treat $A_{r}$, $A_{\theta}$
, $A_{\phi}$and $A_{4}$. Regarding this 4D conformal box, the boundary
conditions for the electromagnetic field are given by

\begin{eqnarray}
\frac{-e}{\hbar}\int_{0}^{R=\infty}A_{r}\cdot dr & = & 2\pi Q_{m}\quad Q_{m}=0,\pm1,\pm2,...\label{gn12-1}\\
\frac{-e}{\hbar}\int_{0}^{2\pi}A_{\theta}\cdot rd\theta & = & 2\pi B_{m}\quad B_{m}=0,\pm1,\pm2,...\label{gn12-2}\\
\frac{-e}{\hbar}\int_{0}^{2\pi}A_{\phi}\cdot r\sin\theta d\phi & = & 2\pi S_{m}\quad S_{m}=0,\pm1,\pm2,...\label{gn12-3}\\
\frac{-e}{\hbar}\int_{0}^{T=\infty}A_{4}\cdot d(ict) & = & 2\pi Q\quad Q=0,\pm1,\pm2,...\label{gn12-4}\end{eqnarray}
 These quantization equations are obvious when thinking about that
the four components of the vector potential are independent from each
other.

The path from the corner $(r=0,\theta=0,\phi=0,t=0)$ in the 4D conformal
box to the corner $(R,\pi,\pi,T)$ corresponds to the half loop of
the big closed loop $L$ as shown in Figure \ref{Fig2}, saying A-B-C.
Considering the half loop A-B-C in Figure \ref{Fig2}, then Eq.(\ref{gn11})
becomes \begin{eqnarray}
\frac{-e}{\hbar}\int_{0}^{R}A_{r}\cdot dr+\frac{-e}{2\hbar}\int_{0}^{2\pi}A_{\theta}\cdot rd\theta+\frac{-e}{2\hbar}\int_{0}^{2\pi}A_{\phi}\cdot r\sin\theta d\phi+\frac{-e}{\hbar}\int_{0}^{T}A_{4}\cdot d(ict) & = & \frac{2\pi n}{2}\label{gn13}\\
n & = & 0,\pm1,\pm2,...\nonumber \end{eqnarray}
 Substituting Eq.(\ref{gn12-1}-\ref{gn12-4}) into Eq.(\ref{gn13}),
we obtain \begin{equation}
Q_{m}+\frac{B_{m}+S_{m}}{2}+Q=\frac{n}{2}\quad n=0,\pm1,\pm2,...\label{gn14}\end{equation}
 We immediately recognize that this equation is the Gell-Mann-Nishijima
relation for hadrons: According to hadron proprieties, we have to
define $Q$ as the charge number, to define $-B_{m}$ as the baryon
number $B$, to define $-B_{m}$ as strangeness number $S$, to define
$\frac{n}{2}$ as the third component of isospin $I_{3}$, typically,
$A_{r}=0$, $Q_{m}=0$, Eq.(\ref{gn14}) becomes the Gell-Mann-Nishijima
formula \begin{equation}
Q-\frac{B+S}{2}=I_{3}\quad I_{3}=0,\pm\frac{1}{2},\pm1,...\label{gn15}\end{equation}
 To note that the charge number $Q$ is conserved, therefore we deduce
that the baryon number $B$ is conserved and the strangeness number
$S$ is conserved too for all hadron reactions. It is important to
realize that the electromagnetic field of a hadron can be regarded
as the higher order mode in terms of its potential-wave-function,
marked by three quantum numbers: charge number $Q$, baryon number
$B$ and strangeness number $S$. This result has three significations:
(1) the formalism we developed about the potential-wave-function is
success and useful for elementary particle classification. (2) the
potential-wave-function provides us a visual model for abstract baryon
number and strangeness number, in analogy with the optical mode $TEM_{qlm}$
in a laser resonator, it provides us an important work platform for
extending our knowledge. (3) the Gell-Mann-Nishijima formula is the
critical condition of the electromagnetic field modes around a hadron.

\section{Charge quantization}

Consider the fourth component quantization Eq.(\ref{gn12-4}) and
the path independence, we have\begin{equation}
\frac{-e}{\hbar}\int_{0}^{T}A_{4}(r)\cdot d(ict)=\frac{-eA_{4}(r)}{\hbar}\int_{0}^{T}d(ict)=\frac{-eA_{4}(r)icT}{\hbar}\label{gn16}\end{equation}

\begin{equation}
A_{4}(r)=\frac{i}{c}V(r)=\frac{i}{c}\int_{r}^{\infty}\mathbf{E}\cdot d\mathbf{l}=\frac{ik_{e}q}{cr}\label{gn17}\end{equation}
 Where we have used the Coulomb's law $E=k_{e}q/r^{2}$, we have \begin{equation}
\frac{-e}{\hbar}\int_{0}^{T}A_{4}(r)\cdot d(ict)=\frac{k_{e}eq}{r}\cdot T\cdot\frac{1}{\hbar}=2\pi Q\quad Q=0,\pm1,\pm2,...\label{gn18}\end{equation}
 From the Heisenberg's uncertainty principle, we immediately recognize
that the potential energy is the uncertain energy $\triangle E=k_{e}eq/r$
for the observed point $x$, and the time limit $T$ for the observed
point $x$ is the uncertain time $\triangle t=T$, both satisfies
the charge quantization and the uncertainty principle, because from
Eq(\ref{gn18}):

\begin{equation}
q=Qe\quad Q=0,\pm1,\pm2,...\label{gn19}\end{equation}
 \begin{equation}
\triangle E\cdot\triangle t=(\frac{k_{e}ee}{r})\cdot T=2\pi\hbar\label{gn20}\end{equation}
 The charge $q$ of the particle $Q$ has been quantized by Eq.(\ref{gn19}).
Here for the first time, the charge quantization can be derived from
the quantum mechanics. In fact, the uncertainty principle Eq.(\ref{gn20})
uses the energy $k_{e}ee/r$ to determine time limit $T$ without
regarding the elementary particle charge number $Q$, it is in minimum
case for the Heisenberg's uncertainty principle, or:\begin{equation}
\triangle E\cdot\triangle t=|\frac{k_{e}eQe}{r}|\cdot T\geq2\pi\hbar\quad Q=\pm1,\pm2,...\label{gn21}\end{equation}
 Where we discard $Q=0$.

\section{Confinement of quarks}

\begin{figure}[htb]
 \centering \includegraphics[bb=150bp 570bp 450bp 760bp,clip]{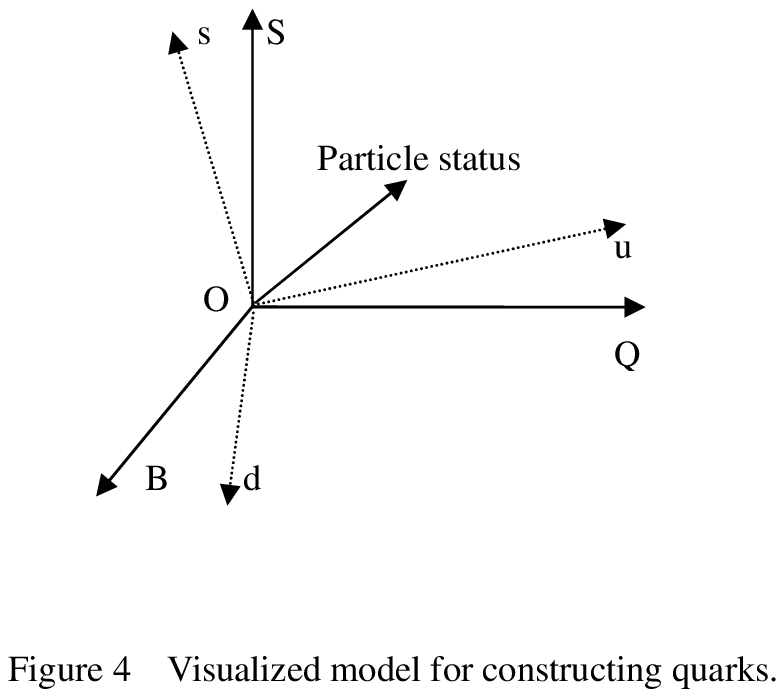}
\caption{visual model for forming quarks.}

\label{Fig4} 
\end{figure}

Quarks are abstract particles; here we focus on how to form the quarks
by using our theory. The charge number $Q$, baryon number $B$ and
strangeness number $S$ are three freedoms for classifying elementary
particles, we can build a frame in which $Q$, $B$ and $S$ are the
three independent axes as shown in Figure \ref{Fig4}. But mathematicians
have rights to rotate the frame $(Q,B,S)$ to a new orientation and
let it becomes a new frame $(u,d,s)$ as shown in Figure \ref{Fig4},
the transformation equations between the old coordinates $(Q,B,S)$
and new coordinates $(u,d,s)$ are given by \begin{equation}
\left[\begin{array}{c}
Q\\
B\\
C\end{array}\right]=\left[\begin{array}{ccc}
R_{11} & R_{12} & R_{13}\\
R_{21} & R_{22} & R_{23}\\
R_{31} & R_{32} & R_{33}\end{array}\right]\cdot\left[\begin{array}{c}
u\\
d\\
s\end{array}\right]\label{gn22}\end{equation}

It is an easy thing to get the transformation matrix $R$ after reading
the basic data of baryons and three quarks, finally it is \begin{equation}
\left[\begin{array}{c}
Q\\
B\\
C\end{array}\right]=\left[\begin{array}{ccc}
2/3 & -1/3 & -1/3\\
1/3 & 1/3 & 1/3\\
0 & 0 & -1\end{array}\right]\cdot\left[\begin{array}{c}
u\\
d\\
s\end{array}\right]\label{gn23}\end{equation}
 Where the axis $u$ represents the up-quark number, the axis $d$
represents the down-quark number; the axis $s$ represents the strange-quark
number. In our viewpoint, the three quark types are another mathematical
representation for the higher order modes of the electromagnetic field
around a particle.

The motivation of this section is to make readers to understand with
reality: (1) The electromagnetic filed modes around a hadron have
three quantum numbers: $Q$, $B$ and $S$, therefore the quark type
number through the mathematical transformation in Eq.(\ref{gn23})
are three. (2) The three quark types have been confined in Eq.(\ref{gn23}),
we have no way to let the three quark types being free particles mathematically
or physically, experiments have no chance to discover free quark in
reality, like virtual phonons in metals.

\section{Six components of electromagnetic field}

Generally speaking, the electromagnetic field around the particle
$Q$ has six component: $(B_{r},B_{\theta},B_{\phi},E_{r},E_{\theta},E_{\phi})$,
we can not naively think that $E_{\theta}=E_{\phi}=0$, specially
the lift time for some hadrons has only $10^{-10}$ seconds or even
shorter. The electrical potential $V$ in the fourth component quantization
Eq.(\ref{gn12-4}) is written as

\begin{equation}
\frac{-e}{\hbar}\int_{0}^{T}A_{4}\cdot d(ict)=\frac{eT}{\hbar}\int_{r}^{\infty}\mathbf{E}\cdot d\mathbf{l}\label{gn24}\end{equation}
 It is also path independent about the electric field, so we have

\begin{equation}
\frac{eT}{\hbar}\oint_{L}\mathbf{E}\cdot d\mathbf{l}=2\pi n\quad n=0,\pm1,\pm2,...\label{gn25}\end{equation}
 Where $L$ is the big close loop linking to the observed point $x$,
as shown in Figure \ref{Fig2}, saying A-B-C-D-A. We are now in a
position to realize: (1) the electric field around the particle has
been quantized by Eq.(\ref{gn25}) mathematically. (2) if the quantum
number $n=0$, then $E_{\theta}=E_{\phi}=0$, then Eq.(\ref{gn25})
is the Ampere's law for the electrostatic field. (3) if $n\neq0$,
then, $E_{\theta}\neq0$ or $E_{\phi}\neq0$, the electric field $\mathbf{E}$
is recognized as a higher order mode ($n$ denotes the mode rank)
with a mathematical singularity at the particle. (3) regarding the
potential-wave-function, the boundary conditions for the electric
field are

\begin{eqnarray}
\frac{eT}{\hbar}\int_{r}^{\infty}E_{r}\cdot dr & = & 2\pi Q\quad Q=0,\pm1,\pm2,...\label{gn26-1}\\
\frac{eT}{\hbar}\int_{0}^{2\pi}E_{\theta}\cdot rd\theta & = & 2\pi B_{e}\quad B_{e}=0,\pm1,\pm2,...\label{gn26-2}\\
\frac{eT}{\hbar}\int_{0}^{2\pi}E_{\phi}\cdot r\sin\theta d\phi & = & 2\pi S_{e}\quad S_{e}=0,\pm1,\pm2,...\label{gn26-3}\end{eqnarray}
 Considering the half loop A-B-C in Figure \ref{Fig2}, then Eq.(\ref{gn12-4})
becomes\begin{eqnarray}
\frac{-e}{\hbar}\int_{0}^{R}A_{r}\cdot dr+\frac{-e}{2\hbar}\int_{0}^{2\pi}A_{\theta}\cdot rd\theta+\frac{-e}{2\hbar}\int_{0}^{2\pi}A_{\phi}\cdot r\sin\theta d\phi\nonumber \\
+\frac{eT}{\hbar}\int_{R}^{\infty}E_{r}\cdot dr+\frac{eT}{2\hbar}\int_{0}^{2\pi}E_{\theta}\cdot rd\theta+\frac{eT}{2\hbar}\int_{0}^{2\pi}E_{\theta}\cdot rd\theta & = & \frac{2\pi n}{2}\label{gn27}\\
n & = & 0,\pm1,\pm2,...\nonumber \end{eqnarray}
 Substituting Eq.(\ref{gn12-1}-\ref{gn12-4}) and Eq.(\ref{gn26-1}-\ref{gn26-3})
into Eq.(\ref{gn27}), we obtain \begin{equation}
Q_{m}+\frac{B_{m}+S_{m}}{2}+Q+\frac{B_{e}+S_{e}}{2}=\frac{n}{2}\quad n=0,\pm1,\pm2,...\label{gn28}\end{equation}
 We immediately recognize that this equation is the Gell-Mann-Nishijima
relation for hadrons.

Therefore, the electromagnetic field mode around a particle has six
quantum numbers: $(Q,B_{e},S_{e},Q_{m},B_{m},S_{m})$, thus the independent
quark type number relating to the six quantum numbers are actually
six. Recalling up to now we have known that the quarks have six types
$(u,d,s,c,b,t)$ and the leptons have six type too, and the electromagnetic
field has six components, we realize that 6 is a magic number for
physics: all physical quantities are sharing the 6 freedoms originating
from the 6 components of electromagnetic field in explicit or implicit
ways\cite{H_Y_Cui}.

Because of the charge conservation, we deduce that these six quantum
numbers are conserved, in agreement with the experiments of hadrons
in conservation law number: six. Of cause, the details of the relationships
concerning quarks seem to be complex and abstract.

\section{Gell-Mann-Nishijima formula in hydrogen atom}

Hydrogen atom consists of a single electron bound to its central nucleus,
a single proton, by the attractive Coulomb force that acts between
them. In the theory of relativistic quantum mechanics, the 4-vector
momentum of an electron $p$ is given by \begin{equation}
(p_{\mu}+qA_{\mu})\psi=-i\hbar\partial_{\mu}\psi\quad\mu=1,2,3,4\label{eq19}\end{equation}
 Where $A$ is the 4-vector potential of the electromagnetic field,
$\psi$ is the wave-function of the electron. We have adopted momentum
$p$ rather than momentum operator $\hat{p}$ in Eq.(\ref{eq19})\cite{H_Y_Cui}.
In our viewpoint, the wave-function $\psi$ is the potential-wave-function
of the atom, because \begin{equation}
\psi=\exp[\frac{i}{\hbar}\int_{l}(p_{\mu}+qA_{\mu})\cdot dx_{\mu}]\label{eq20}\end{equation}
 Where $l$ is an arbitrary path from the observed point x to the
infinity. In analogy with the electric field modes discussed in the
preceding sections, for any closed loop $L$ containing the nucleus
in Figure \ref{Fig1}, the wave function $\psi$ for the electron
in a hydrogen atom is quantized by \begin{equation}
\frac{1}{\hbar}\oint_{L}(p_{\mu}+qA_{\mu})\cdot dx_{\mu}=2\pi n\quad n=0,\pm1,\pm2,...\label{eq21}\end{equation}
 The nucleus of the hydrogen atom provides the 4-vector potential
for the electron: \begin{equation}
A_{r}=A_{\theta}=A_{\phi}=0;\quad A_{4}=\frac{ie}{cr}\label{eq22}\end{equation}
 Where we take $k_{e}=1$ in the Coulomb's law: Gaussian units. Because
the electron keeps the angular momentum conservation, the angular
momentum magnitude $J=r\sqrt{p_{\theta}^{2}+p_{\phi}^{2}}$ is a constant
and its z-axis component magnitude $J_{z}=r\sin\theta p_{\phi}$ is
a constant too. The energy of the electron $E$ also is a constant
for its stationary state:$\psi=\psi(\mathbf{r})e^{-iEt/\hbar}$, thus\begin{eqnarray}
\exp(-iEt/\hbar) & = & \exp[\frac{i}{\hbar}\int_{l}(p_{4}+qA_{4})\cdot dx_{4}]\label{eq231}\\
p_{4} & = & \frac{-E-e^{2}/r}{ic}\qquad x_{4}=ict\label{eq232}\end{eqnarray}
 From $p_{\mu}p_{\mu}=-m_{e}^{2}c^{2}$, where $m_{e}$ is the mass
of the electron, we obtain \begin{eqnarray}
p_{r} & = & \sqrt{-m_{e}^{2}c^{2}-p_{\theta}^{2}-p_{\phi}^{2}-p_{4}^{2}}=\sqrt{-m_{e}^{2}c^{2}-\frac{J^{2}}{r^{2}}+\frac{1}{c^{2}}(E+\frac{e^{2}}{r})^{2}}\label{eq241}\\
p_{\theta} & = & \sqrt{(p_{\theta}^{2}+p_{\phi}^{2})-p_{\phi}^{2}}=\frac{1}{r}\sqrt{J^{2}-\frac{J_{z}^{2}}{\sin^{2}\theta}}\label{eq242}\end{eqnarray}
 Therefore, quantization Eq.(\ref{eq21}) for the loop in Figure \ref{Fig2}
becomes \begin{eqnarray}
\frac{1}{k}\oint_{L}p_{r}\cdot dr+\frac{1}{k}\oint_{L}p_{\theta}\cdot rd\theta+\frac{1}{k}\oint_{L}p_{\phi}\cdot r\sin\theta d\phi & = & 2\pi n\label{eq25}\\
n & = & 0,\pm1,\pm2,...\nonumber \end{eqnarray}
 Where the loop integral about the time $t$ component vanishes for
the constant energy $E$ in Eq.(\ref{eq231}), while the loop becomes
three quantization equations\begin{eqnarray}
\frac{1}{\hbar}\int_{0}^{\infty}p_{r}\cdot dr & = & \pi f\quad f=0,\pm1,\pm2,...\label{eq261}\\
\frac{1}{\hbar}\int_{0}^{2\pi}p_{\theta}\cdot rd\theta & = & 2\pi g\quad g=0,\pm1,\pm2,...\label{eq262}\\
\frac{1}{\hbar}\int_{0}^{2\pi}p_{\phi}\cdot r\sin\theta d\phi & = & 2\pi m\quad m=0,\pm1,\pm2,...\label{eq263}\end{eqnarray}
 They also can be written as \begin{eqnarray}
\frac{1}{\hbar}\int_{0}^{\infty}\sqrt{-m_{e}^{2}c^{2}-\frac{J^{2}}{r^{2}}+\frac{1}{c^{2}}(E+\frac{e^{2}}{r})^{2}}\cdot dr & = & \pi f\quad f=0,\pm1,\pm2,...\label{eq271}\\
\frac{1}{\hbar}\int_{0}^{2\pi}\sqrt{J^{2}-\frac{J_{z}^{2}}{\sin^{2}\theta}}\cdot d\theta & = & 2\pi g\quad g=0,\pm1,\pm2,...\label{eq272}\\
\frac{1}{\hbar}\int_{0}^{2\pi}J_{z}d\phi & = & 2\pi m\quad m=0,\pm1,\pm2,...\label{eq273}\end{eqnarray}
 These definite integrals have been evaluated on the ranges: these
integrands take real values in the previous paper\cite{H_Y_Cui},
as the result, they are \begin{eqnarray}
\frac{\pi E\alpha}{\sqrt{m_{e}^{2}c^{4}-E^{2}}}-\pi\sqrt{(g+|m|)^{2}-\alpha^{2}} & = & \pi f\quad f=0,\pm1,\pm2,...\label{eq281}\\
\frac{2\pi}{\hbar}(J-|J_{z}|) & = & 2\pi g\quad g=0,\pm1,\pm2,...\label{eq282}\\
\frac{2\pi}{\hbar}J_{z} & = & 2\pi m\quad m=0,\pm1,\pm2,...\label{eq283}\end{eqnarray}
 Where the $\alpha=e^{2}/(\hbar c)$ is known as the fine structure
constant. It is very easy to get the energy levels of the electron
from the Eq.(\ref{eq281}): \begin{equation}
E=m_{e}c^{2}\left[1+\frac{\alpha^{2}}{(\sqrt{(g+|m|)^{2}-\alpha^{2}}+f)^{2}}\right]^{-\frac{1}{2}}\label{eq29}\end{equation}
 This result, Eq.(\ref{eq29}), is completely the same as the calculation
of Dirac wave equation for hydrogen atom, it is just the fine structure
of hydrogen atom energy.

Considering the half loop A-B-C in Figure \ref{Fig2} about the nucleus,
then Eq.(\ref{eq25}) becomes \begin{eqnarray}
\frac{1}{k}\int_{0}^{\infty}p_{r}\cdot dr+\frac{1}{2k}\int_{0}^{2\pi}p_{\theta}\cdot rd\theta+\frac{1}{2k}\int_{0}^{2\pi}p_{\phi}\cdot r\sin\theta d\phi & = & \frac{2\pi n}{2}\label{eq30}\\
n & = & 0,\pm1,\pm2,...\nonumber \end{eqnarray}
 Substituting Eq.(\ref{eq261}-\ref{eq263}) into Eq.(\ref{eq30}),
we find that the Gell-Mann-Nishijima formula also exists in hydrogen
atom: \begin{equation}
\frac{f}{2}+\frac{g+m}{2}=\frac{n}{2}\quad n=0,\pm1,\pm2,..\label{eq31}\end{equation}
 In the stationary state spectrum of hydrogen atom, frequently $f=0$.
This Gell-Mann-Nishijima formula is the critical condition of the
wave function modes around a hydrogen atom.

Among Eq.(\ref{eq261}-\ref{eq263}), it is unfair to $\psi(r)$,
$\psi(\theta)$ and $\psi(\phi)$, because we let $\psi(r)$ form
a standing wave (half integral multiple) from $r=0$ to $r=\infty$
while $\psi(\theta)$ and $\psi(\phi)$ are simple wave (integral
multiple). As we have mentioned, we must fairly treat the three components
in their conformal box (Figure \ref{Fig3}). Therefore, Fairly, if
we let they all are integral multiple, then their quantum numbers
satisfy the standard Gell-Mann-Nishijima formula.

\section{Visual Model of hadrons}

\begin{figure}[htb]
 \centering \includegraphics[bb=120bp 590bp 420bp 760bp,clip]{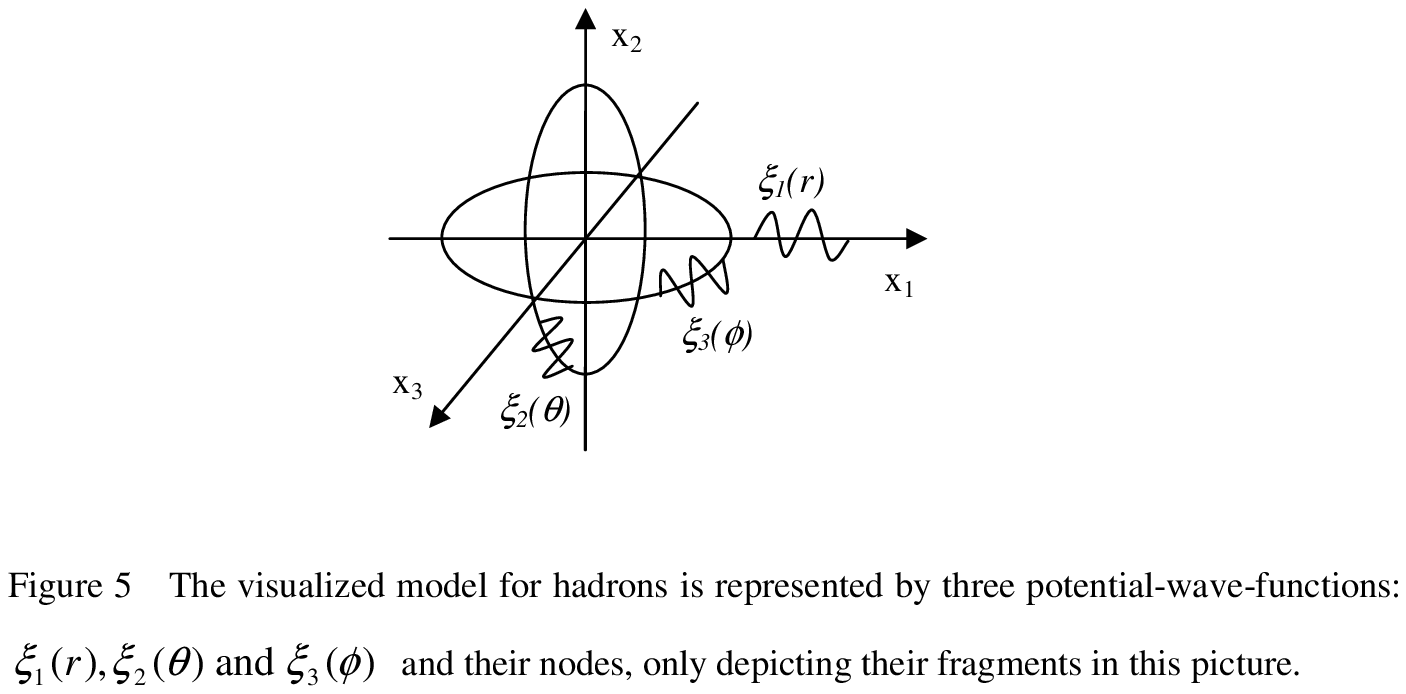}
\caption{The visual model for hadrons is represented by three potential-wave-functions:
$\xi_{1}(r),\xi_{2}(\theta),\xi_{3}(\phi)$ and their nodes, where
only depicting their fragments.}

\label{Fig5} 
\end{figure}

Consider a baryon of charge number $Q$, Baryon number $B$ and strangeness
number $S$, its potential-wave-function can be separated into three
parts as \begin{equation}
\xi(\mathbf{r})=\xi_{1}(r)\xi_{2}(\theta)\xi_{3}(\phi)\label{eq32}\end{equation}
 Where $r$ varies in the range $(0,\infty)$ with $Q+1$ nodes, $\theta$
varies in the range $(0,2\pi)$ with $2B+1$ nodes, $\phi$ varies
in the range $(0,2\pi)$ with $2S+1$ nodes, as shown in Figure \ref{Fig5},
where only depicting each fragments for them. The picture in Figure
\ref{Fig5} is the visual model for hadrons.

According to Eq.(\ref{gn23}), the visual model for quarks can established
from \begin{equation}
\left[\begin{array}{c}
u\\
d\\
s\end{array}\right]=\left[\begin{array}{ccc}
2/3 & -1/3 & -1/3\\
1/3 & 1/3 & 1/3\\
0 & 0 & -1\end{array}\right]^{-1}\cdot\left[\begin{array}{c}
Q\\
B\\
S\end{array}\right]\label{eq33}\end{equation}

\section{Conclusions}

When an electron probes another elementary particle Q, the wave function
of the electron can be separated into two independent parts, the first
part represents the electronic motion, the second part represents
the electromagnetic field mode around the particle Q. In analogy with
optical modes $TEM_{nlm}$ for a laser resonator, when the electromagnetic
field around the particle Q forms into a mode, the quantum numbers
of the mode satisfy the Gell-Mann-Nishijima formula, these quantum
numbers are recognized as the charge number, baryon number and strangeness
number. As a result, it is found that the Gell-Mann-Nishijima formula
is the critical condition of the electromagnetic field modes around
a hadron. It is important to realize that the electromagnetic field
modes around a hadron provide us a visual model to understand the
abstract baryon number $B$ and strangeness number $S$ in reality.

\end{document}